# Flame spread over thin circular PMMA rods

Manu B V[a*], Amit Kumar[a].

a. National Centre for Combustion Research and Development and Department of Aerospace Engineering, Indian Institute of Technology Madras, Chennai, India.

* Corresponding author.

E-mail address: ae19d017@smail.iitm.ac.in (Manu B V).

## Abstract

In the present article, series of opposed flow, downward flame spread experiments are conducted using 80 mm long cast PMMA cylinders of 1 mm and 0.5 mm diameters at normal and micro gravity environments. All the experiments are conducted at 21 % of oxygen by volume in nitrogen at 1 atmospheric pressure with varying the opposed flow speed from 0 cm/s to 25 cm/s. Flame spread rate, flame shape, preheating, and pyrolysis parameters are observed and analysed by tracking the leading edge of the flame by digital image processing, measuring flame field temperatures, schileren imaging and microscopic observation of burnt solid cones. It is concluded from the study that, flame spreads faster in microgravity for all diameters and opposed flows considered in the study, because of longer preheating length which is confirmed by temperature measurement using 50 microns thermocouples in a flame field. The thinner the fuel faster the flame spread rate because of curvature effect and reduced thermal inertia. In normal gravity, the flame spread rate reduces very slightly with increasing opposed flow while in micro gravity it decreases monotonically for both the diameter of fuels considered for the study.

## Introduction

The fire safety in microgravity is extremely important concern for ensuring the space missions safety. Especially in case of missions involves human crews and other living beings. The spacecraft and its habitat involves various thin cylindrical materials like electrical wires, optical cables, tying cables, astronauts hair, etc. Hence it is essential to have a combustion and fire safety experiments in spacecraft environment. Another usefulness of these experiments is, they will serve for validating the numerical and theoretical models. Many experimental and numerical studies are done on electrical wires thick polymer cylinders. In normal gravity, the flame spread over cast clear PMMA rods of diameter 1.6 mm to 12 mm will increase as the angle changes from vertically downward configuration. From energy balance analysis by measuring detailed temperature field in flame and solid it is observed that for a cylindrical fuels the preheat length will be in the order of 4 mm to 5 mm and the gas phase conduction and conduction inside the solid are only responsible for flame spread while the gas phase radiation is negligible with solid re-radiation accounting for the loss [1]. In a same way, in quiescent microgravity environment the flame will extinguish over a flat fuel even at higher oxygen level of 50 % because of surface re-radiation but the flame spreads steadily over cylindirical samples of 2 mm and 6.4 mm diameter because of no importance of surface re-radiation in cylindrical fuel[2]. In microgravity the black PMMA cylinders of diameter 6.4 mm and above will produce a smoothly spreading spherical flame which shows the increasing and decreasing opposed flame spread trend with increasing the flow from 0.4 cm/s to 8 cm/s in a oxygen level of around 21% [3]. At high oxygen level of 49.4 %, 10 mm cast clear PMMA rod will produce a bluish spherical flame with frequent bursting because of high heat flux at high oxygen availability. It show a increasing trend of flame spread rate at lower opposed flow speeds of up to 12 cm/s [4]. In hyperoxic quiescent microgravity environment thin cast clear PMMA cylinders of diameters 1.5 mm to 4 mm burns with random bursting and shooting at leading edge. The flame shape is of mushroom shape with longer trailing edge and grows as oxygen concentration increases [5]. In case of ETFE wire combustion in microgravity, the radiation heat loss from the flame is very less hence less temperature drop of flame compare to normal gravity flame. The gas phase temperature field obtained by thermocouples reveals the thicker longer preheat length in microgravity and hence higher spread rate [6].

**Fuels**

The samples used for experiments are cast clear PMMA cylinders of 80 mm long and 1 and 0.5 mm in diameter. Thin PMMA cylinders are cast in a lab instead of conducting experiments using available extruded PMMA fibres to study the flame spread without the effect of melting and dripping which is the nature of extruded PMMA burning at normal gravity. The cylinders are produced by the polymerization of an MMA monomer using a BP polymerizing agent. The monomer and polymerizing agents of a 1:0.1 ratio of weight are activated by heating at 90 degrees Celsius in an hot air oven. The viscous activated liquid is filled into melting point glass capillaries of 100 mm long and 1and 0.5 mm inner diameters. The filled capillaries are cured at 33 degree Celsius for the first 12 hours, 36 degree Celsius for the next 36 hours, and 50 degree Celsius for the next 48 hours in a hot water bath as shown in Fig.1. A forced convection of water is facilitated in the water bath to enhance the heat dissipation to avoid bubble formation due to polymerization. The heating process is selected based on trial and error experiments conducted to obtain visible bubble-free cylinders.

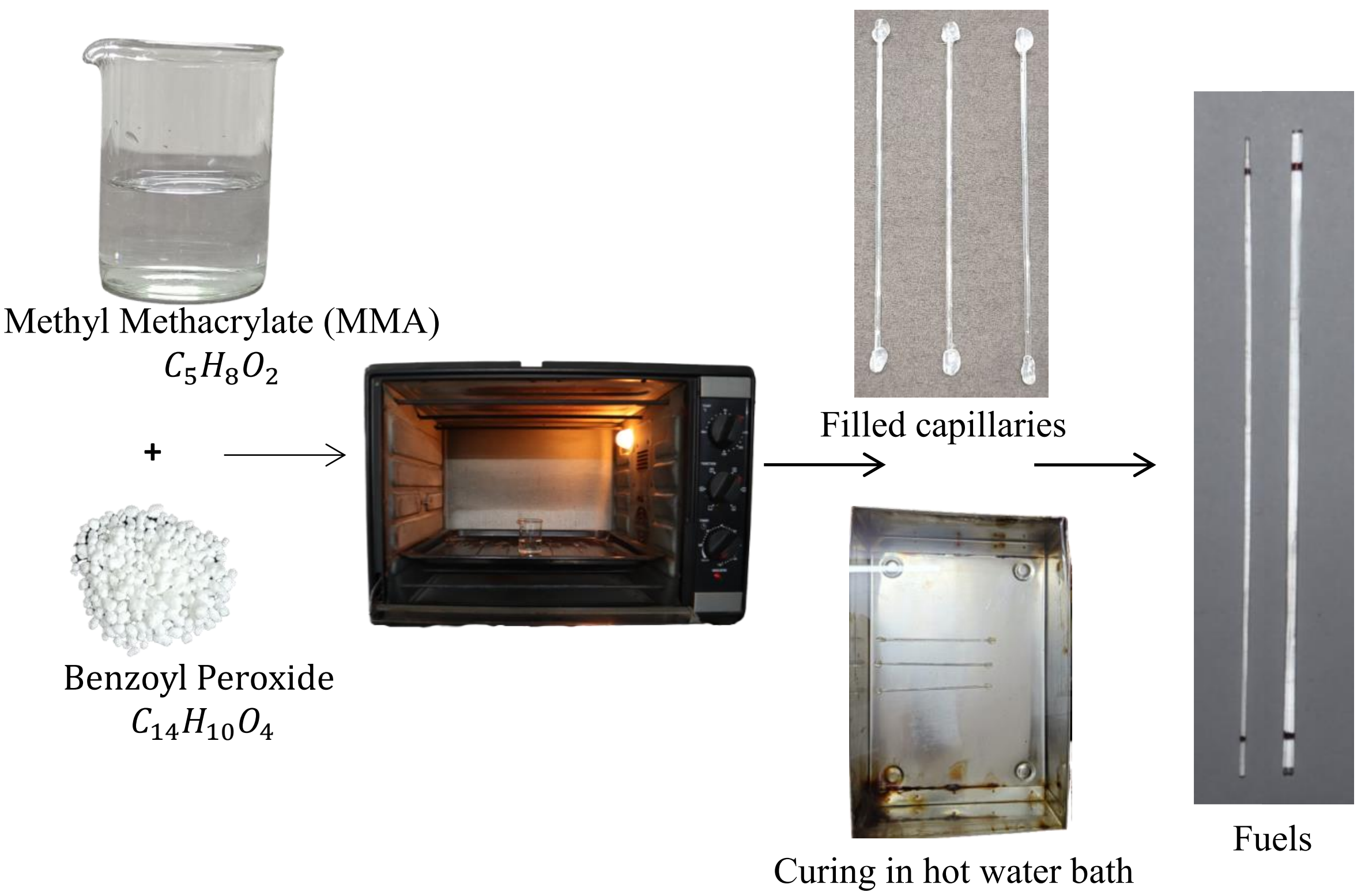

**Figure 1.** Casting process of PMMA cylinders

## Experiments

The normal gravity experiments are conducted at earth's gravity condition. The experimental setup is consisting of combustion tunnel, enclosure, cameras, mountings, data acquisition and control systems as shown in Fig. 2. The combustion tunnel is a rectangular duct of 12 cm X 12cm cross-section and 27cm height. The lower part comprises metallic stages for accommodating a fan and flow conditioning unit (honeycomb and meshes), and the upper part is the test section made up of transparent polycarbonate plates. The opposed flow speed is varied by controlling a supply voltage to the DC fan and flow is characterised at the plane located at the centre of the test section, by using a one directional hot wire anemometer. The sample holder is a thin capillary tube of 1mm inner diameter and 1.2mm outer diameter attached to a thin stainless steel strip to hold the fuel vertically upwards with minimum flow disturbance. A nichrome coil is fixed at the topmost of the chamber to ignite the fuel at the top end. The complete combustion tunnel is enclosed within airtight outer chamber of 44 litre volume to maintain an environment of required oxygen concentration. The combustion tunnel is facilitated with a necessary mountings for placing a digital camera to capture a flame spread process. A canon M50 Mark 2 with 50 FPS camera is used to capture a flame spread. The spread rate is calculated by tracking a leading edge of a flame by using a image processing algorithm called Flame Spread Analyser developed in MATLAB.

The microgravity flame spread experiments are conducted at 2.5 s drop tower and it is shown in Fig. 3. The combustion chamber used in microgravity experiments is same as used in normal gravity experiments.

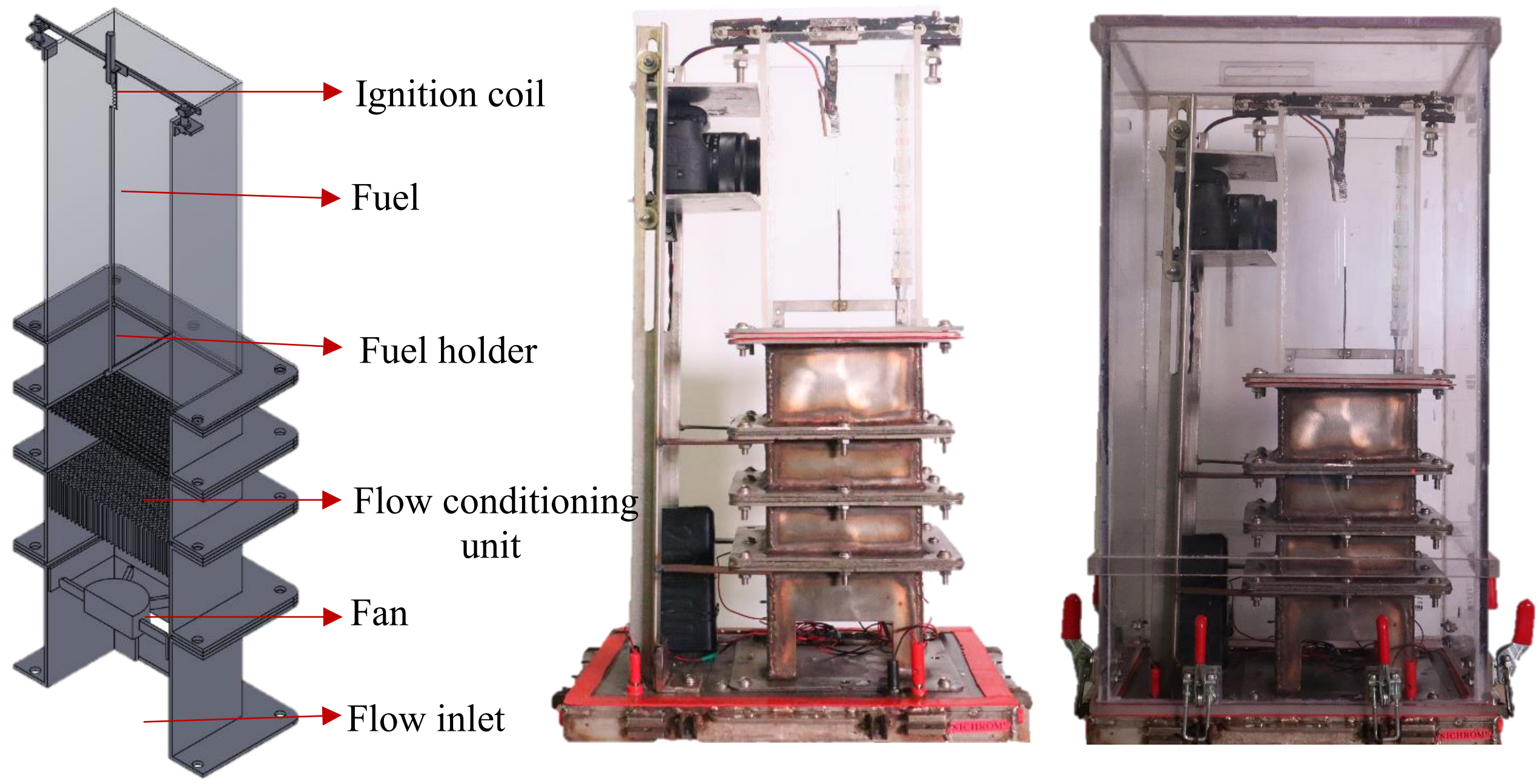

**Figure 2(a):** Schematic of combustion tunnel **(b):** Photograph of combustion tunnel **(c):** Photograph of combustion chamber

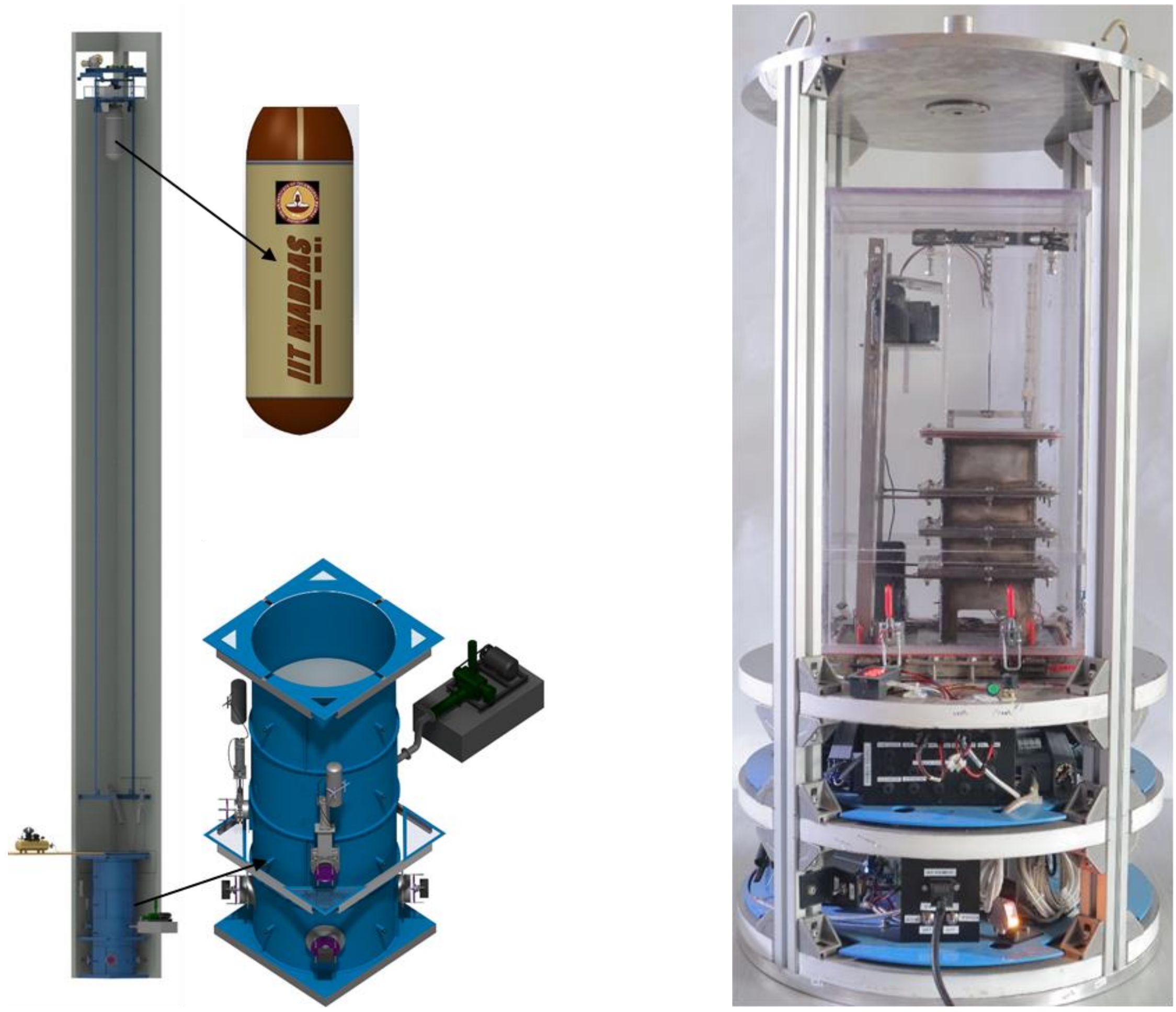

**Figure 3: (a) Schematic of drop tower (b) Photograph of drop capsule**

## Flame shape and behaviour

Figure 4 shows instantaneous frames of flame for different diameters, flow speeds, and gravity conditions. The normal gravity flame is a smoothly spreading intense orange coloured with bluish leading edge. The flame is more spherical in shape for the case of 0.5 mm cylinders and it is slightly lifted vertically upwards in case of 1 mm cylinders. As the flow speed increases it pushes the flame towards the surface of solid and cools down the flame. Hence as the flow increases flame becomes thinner and shorter. In microgravity a unique orange coloured mushroom shaped flame with random bursting and shooting at leading edge is observed. This bursting and shooting of leading edge will get calmer as flow speed increases because of cooling of flame and opposition of shooting by pushing effect of flow.

The random bursting and shooting at leading edge in microgravity flame is clearly shown in top and bottom view of the flame in Figure 4(b). In general the PMMA burns with bursting because of intense heating of hard PMMA at leading edge and its intensity is depends on magnitude of heat flux. Only the leading edge of the flame will propagates over a hard PMMA and as it passes the solid will get soft by heating. The hard part of solid will store high amount of heat and it cracks and burst once the threshold is crossed while molten soft part of solid boils and bubbles smoothly. This phenomenon is not clearly distinguished in downward burning flame because, in a steady downward burning flame the bursting zone and boiling zones are very short in that also the bursting zone is still shorter compared to boiling zone.

Hence a upward burning configuration where bursting and boiling zones are longer enough is selected for schlieren experiments. The Figure 5. is a sequence of schlieren photographs of upward burning process of 1 mm diameter PMMA at 21% oxygen at normal gravity condition. The zone where flame leading edge heats the hard PMMA is a burst zone and the zone of molten PMMA is a boiling zone. It is clearly seen that the bursting zone moves with flame leading edge by leaving the boiling zone behind. At final flame travels completely over a hard part of solid fuel hence complete solid becomes soft hence there will be no bursting and shooting phenomenon is observed and fuel vapour comes smoothly and forms a calm flame.

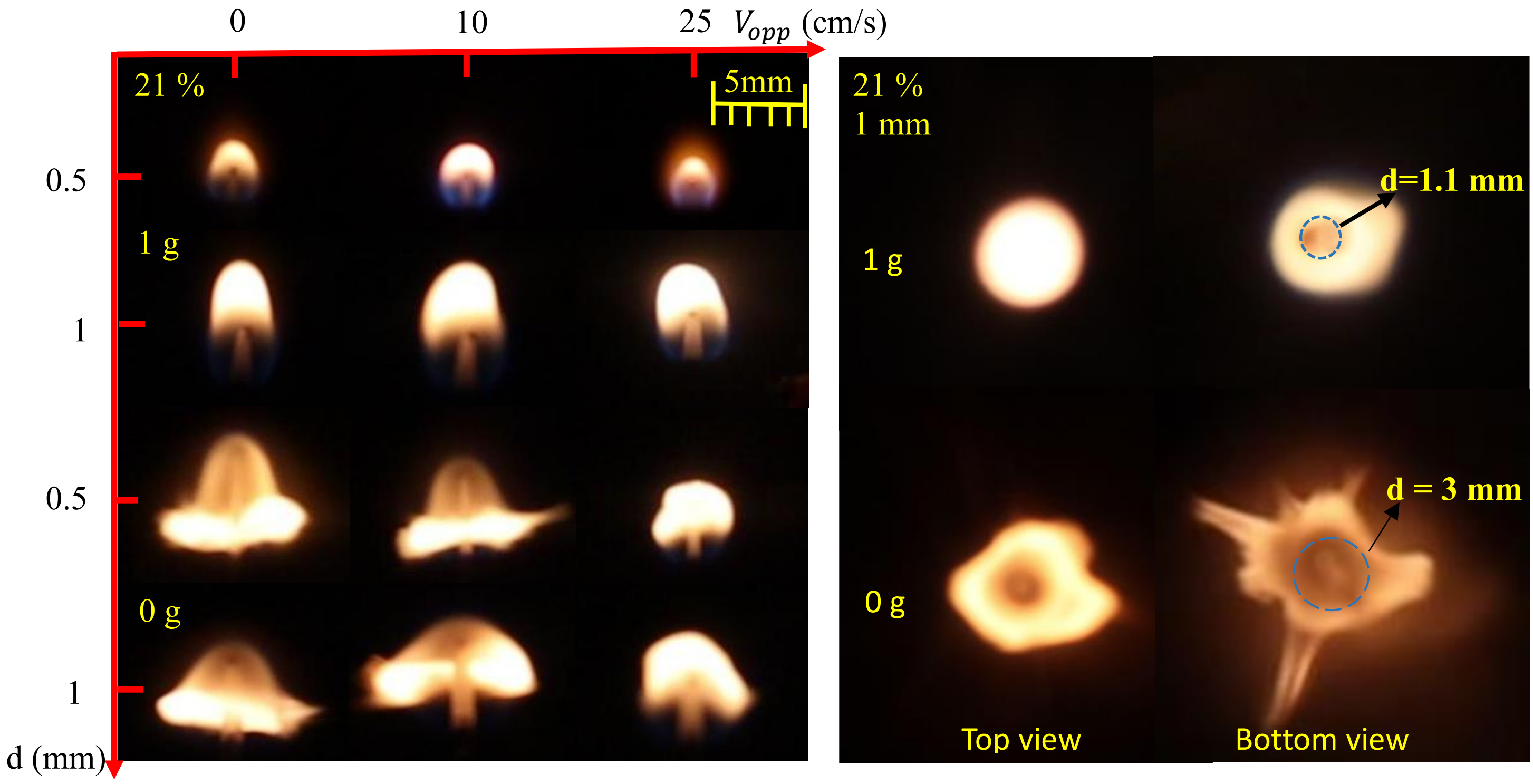

**Figure 4: (a) Instantaneous photographs of flame (b) Bottom and top view of flame**

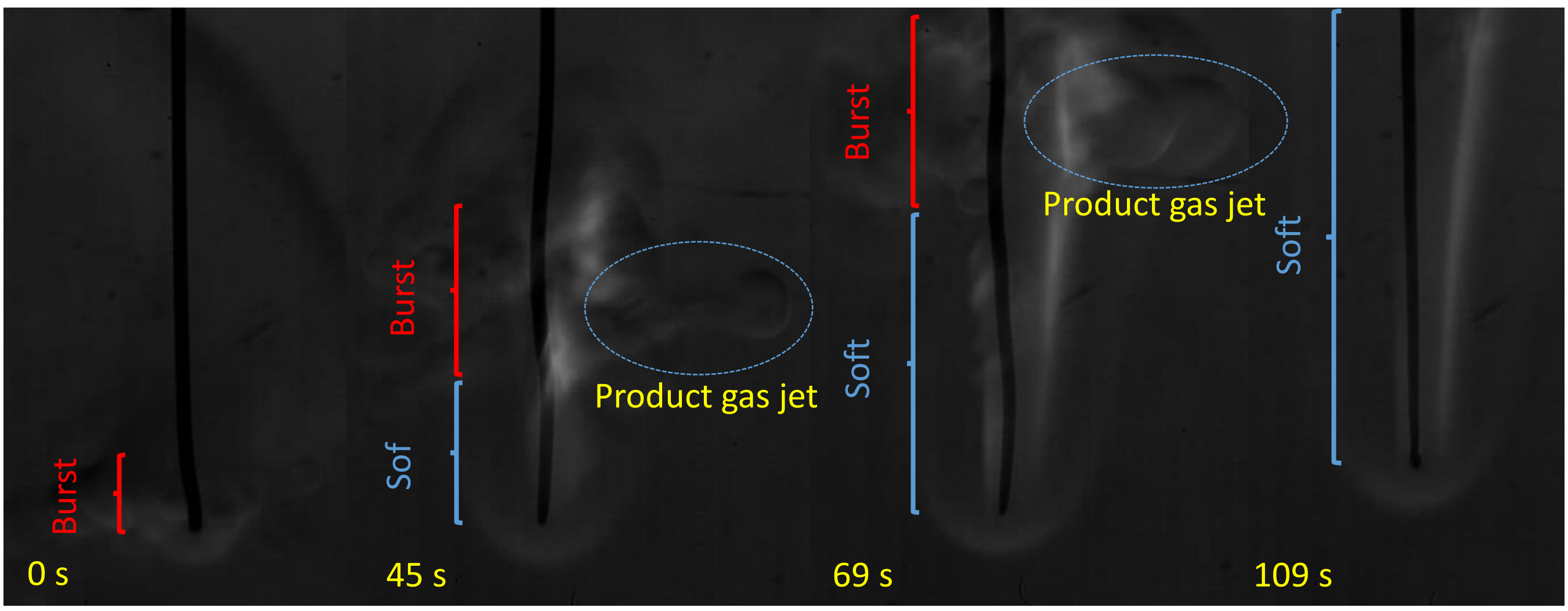

**Figure 5: Schlieren images of upward flame spreading**

In normal gravity flame spreads smoothly without shooting because, the buoyancy induced flow will push the flame inside so it opposes the jets coming out of flame to some extent as explained in Figure 6. Another reason is there will be enough oxygen transport with in the flame zone. Hence the fuel vapour comes out as jet will react with in the flame sheet and only product gasses will come out as jets which are capture in schlieren imaging.

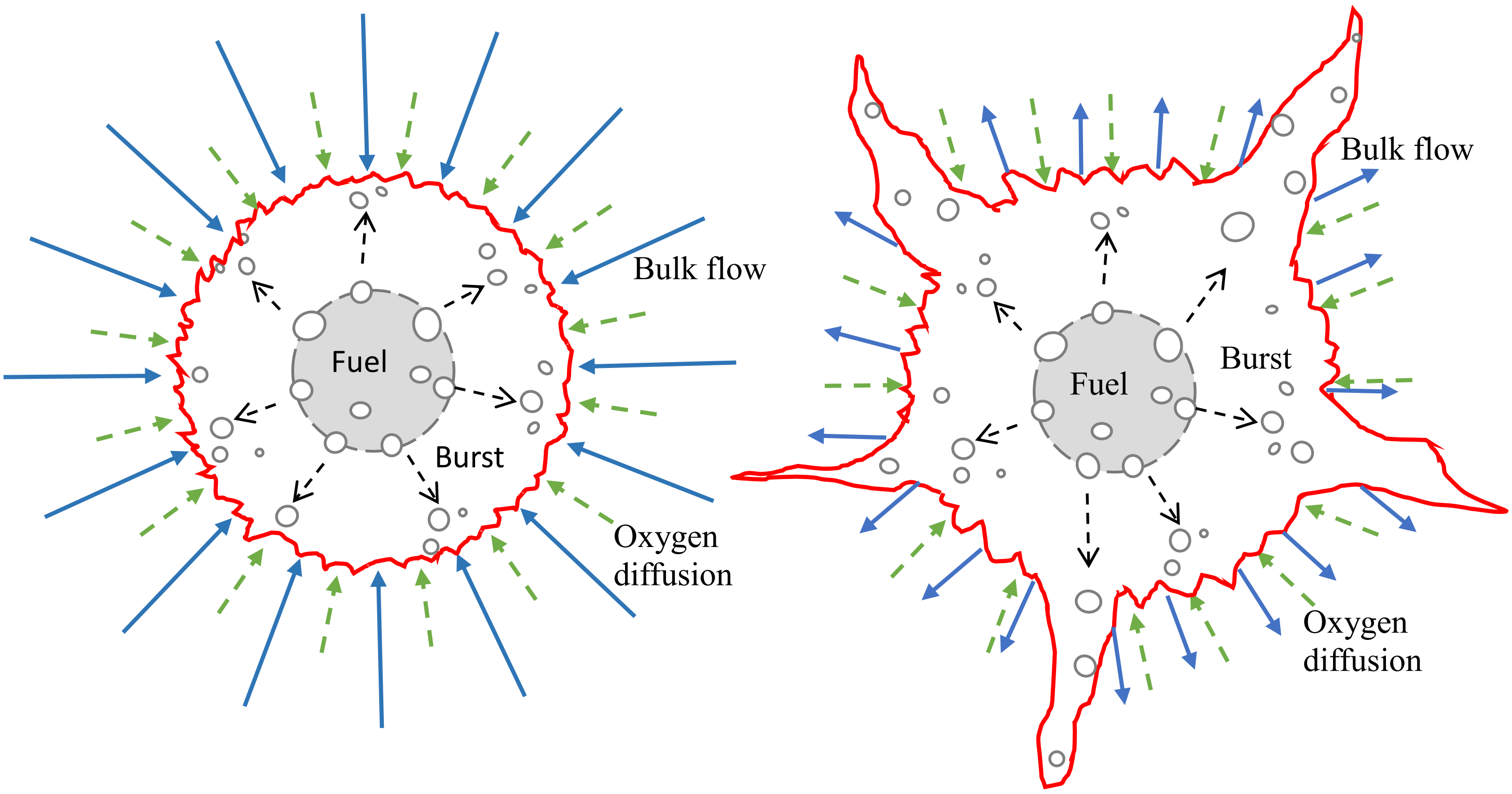

**Figure 6:** PMMA burning and flame shooting process in (a) normal gravity (b) microgravity

In microgravity the flame acts as a bluff body hence it pushes the flow outwards which offers no opposition to jets coming out of fuel by bursting. Diffusion and forced convections are the mode of oxygen transport in microgravity therefore there will be limited oxygen available within the flame zone because of which fuel vapour shoots out as jet and reacts with oxygen available out side of flame zone which are appears as flame shootings. As the forced flow increases the oxygen transport to the flame zone increases and it also offers some opposition to jets coming out of flame because of which the flame becomes smoother as opposed flow increases which is clearly shown in instantaneous photographs of flame spread in microgravity.

The intense bursting phenomenon happens only whenever the incident heat flux to preheat region crosses a particular threshold value. For a particular diameter of the fuel the threshold will reach by increasing the oxygen level or at particular oxygen level the incident heat flux will rise exponentially as the diameter of the fuel reduces because of curvature effect which is explained in later section. Fuels of thinner than this diameter corresponding to threshold heat flux will always burns with bursting at this level of oxygen.

**Flame spread rate**

The Fig.4 shows the flame spread rate variation with opposed flow. Here all the spread rates are average of 2 e xperiments repeated at same experimental conditions. In normal gravity, the flame spread rate for the 1 mm diameter fuel is 0.68 mm/s at no opposed flow and it slightly reduces as the flow increases and it is 0.53 mm/s for 25 cm/s opposed flow. A similar trend is observed in case of 0.5 mm diameter fuels also. This almost constant flame spread rate with increasing opposed flow is due to buoyancy induced upward convection of order 25 to 30 cm/s for unit heated length in vertical direction which is of grater magnitude compare to the opposed flows considered in the study. A monotonically decreasing trend of spread rate is observed in microgravity for both 1 mm and 0.5 mm diameter fuels. It is observed because in the selected oxygen level the flow has only cooling effect on flame. It means even at quiescent microgravity the flame is getting enough oxygen for well established structure and spread. Hence as the opposed flow increases the flame will gets cooler and hence lower the heat feedback to preheat region and lower the spread rate.

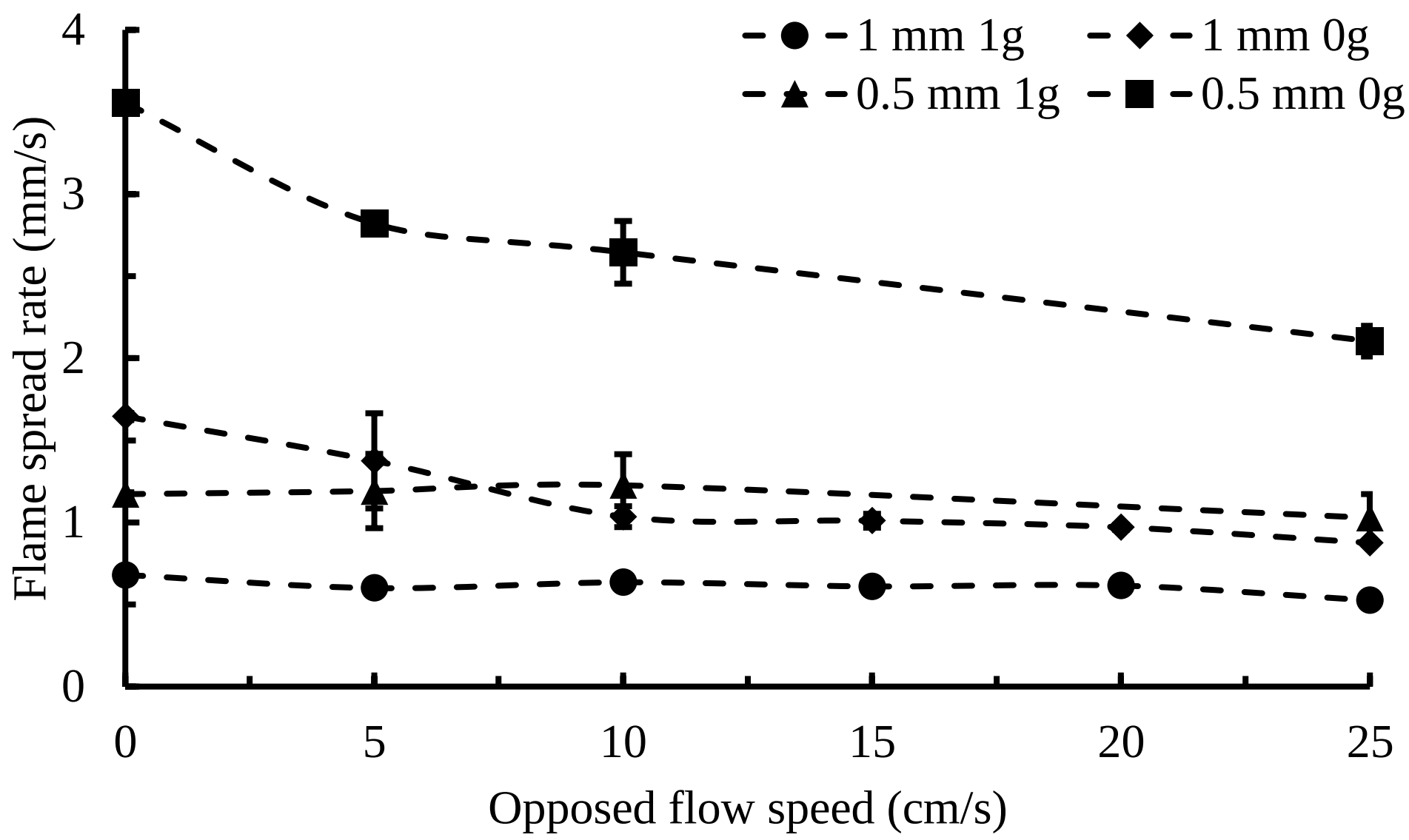

**Figure 4.** Flame spread rate variation with flow

The thinner the fuel faster the spread rate because of the reduced thermal inertia and enhanced heat flux to preheat region due to curvature effect as the size of cylinders reduces as explained in Fig.5.

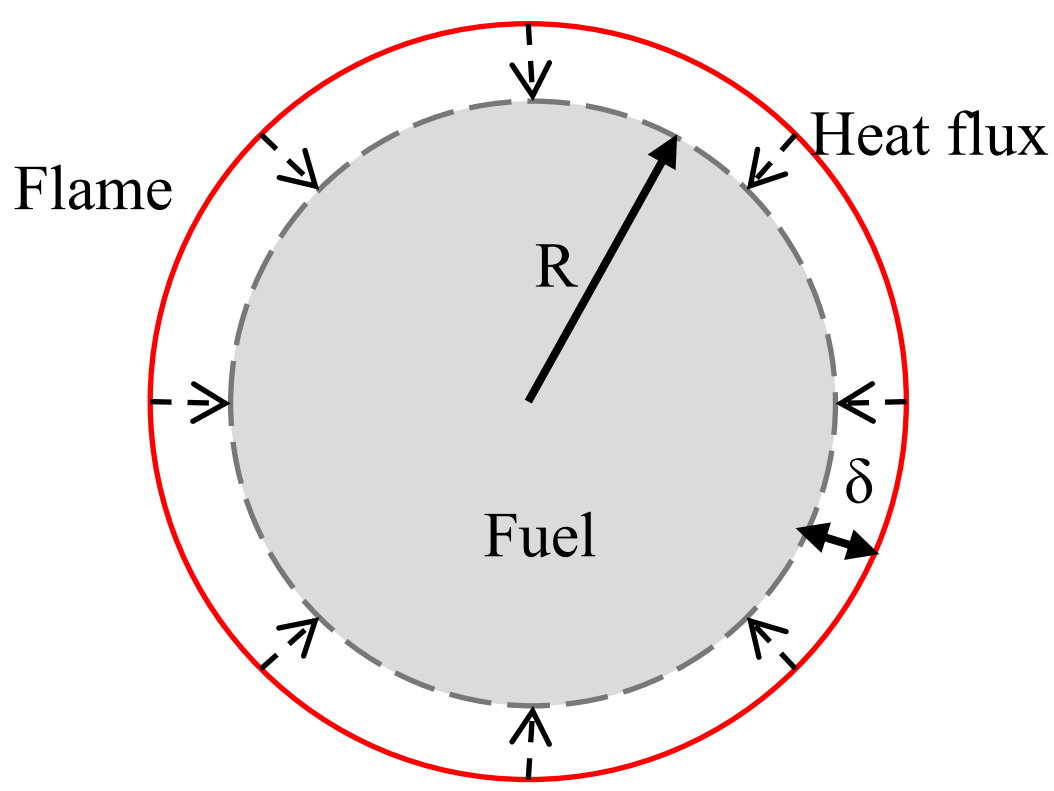

Solid surface area $A_s = 2\Pi RL$

Flame surface area $A_F = 2\Pi(R+\delta)L$

Area ratio A = $\frac{A_F}{A_s}$

**Figure 5.** Schematic of effect of curvature in cylindrical fuels

The ratio of flame area ($A_F$) to fuel surface area ($A_s$) called area ratio (A) is always greater than 1 for a unit length of a cylinders. For a fixed flame standoff distance (δ) It increases exponentially with decreasing diameter of cylinders. Hence, the heat feedback from the flame to preheat region increases exponentially with decreasing diameter for a fixed flow and oxygen level. Hence, as the diameter of the cylinders reduces flame spread rate increases.

**Conclusions**

In the present work, opposed flow downward flame spread experiments are conducted on cast circular PMMA rods of 0.5 mm and 1 mm diameters in both normal gravity and microgravity environments. Flame spread, shape, and behaviour are studied at ambient (21%) and elevated ( 23%, and 40%) oxygen concentrations in both normal gravity and microgravity environments at various external opposed flow speeds ranging from 0 cm/s to 25 cm/s. Limited set of experiments are also conducted in normal gravity for no external flow at various oxygen concentrations varying from 21% to 60%. The major findings from the study are as follows,

1) The present experimental work contributes to the flame spread data for cast PMMA in normal gravity and microgravity environments for the rods of smallest diameter (1 mm and 0.5 mm) so far in thermally regime. Such data have are difficult to obtain due to fabrication challenges.

2) At low oxygen levels (here 21% and 23%) in normal gravity environment the spreading flame is laminar. While in the microgravity environment, the flame acquires the shape of a mushroom with distorted leading edge because of sporadic fuel jets emanating from the bursting bubbles in the solid fuel. This mushroom shape flame is very different from the round shape flame reported in past works on large diameter cylindrical PMMA rods.

3) At low oxygen levels (here 21% and 23%), for both fuel rod diameters of 0.5 mm and 1 mm the flame spreads faster in microgravity at all external flow speeds compared to corresponding condition in normal gravity. This is due to longer preheat length in microgravity. Further, the flame spread rate decreases with increase in the external opposed flow speed in both normal gravity and micro gravity environments. This trend is different from the non-monotonic flame spread variation with flow speed reported for thicker PMMA rods. The decrease of flame spread rate with external flow speed is steeper in microgravity. At high oxygen levels, (here 40%) the flame spread rates in normal gravity and microgravity are nearly same at all external flow speeds.

4) The sectional view of partially burnt 1 mm PMMA rod under microscope showed bubbles all over the pyrolysis and preheat segments. Numerous small bubbles near the surface and fewer larger bubbles within the solid are noted. The bubbles formed in microgravity appear bigger (50-280 µm) compared to those formed in normal gravity (40-180 µm). The measured solid cone half angle (or regression angle) of the pyrolysis segment is about 7° in normal gravity and about 1.5° in microgravity.

5) The preheat segment length for spreading flame over 1 mm PMMA rod at 21% oxygen and no external flow condition was determined by temperature measurement using fine thermocouples. The preheat length in microgravity is found to be longer (9.82 mm) compared to the preheat length of normal gravity (3.44 mm). It is shown here that reasonable theoretical estimate of this preheat length, especially in quiescent microgravity requires Stefan velocity to be included in defining the reference velocity.